\documentclass[aps, english, superscriptaddress]{revtex4}
\usepackage[T1]{fontenc}
\usepackage[latin9]{inputenc}
\usepackage{babel}
\usepackage{float}
\usepackage{amsmath}
\usepackage{amssymb}
\usepackage{graphicx}
\usepackage[unicode=true]
 {hyperref}
\usepackage{breakurl}
\usepackage{graphics}
\usepackage{xcolor}

\makeatletter

\usepackage{amsmath}
\usepackage{titlesec}
\usepackage[below]{placeins}
\titlespacing*{\subsubsection}
{0pt}{1.5\baselineskip}{1\baselineskip}
\titlespacing*{\section}
{0pt}{2\baselineskip}{1\baselineskip}

\makeatother

\begin{document}

\title{Quantum-Optical set-up for the Monty Hall problem}

\author{L. F. Quezada}
\email{luis.fernando@correo.nucleares.unam.mx}
\affiliation{Instituto de Ciencias Nucleares, Universidad Nacional Aut\'{o}noma de M\'{e}xico, 04510 Ciudad de M\'{e}xico, M\'{e}xico}
\affiliation{Centro de Ciencias de la Complejidad, Universidad Nacional Aut\'{o}noma de M\'{e}xico, 04510 Ciudad de M\'{e}xico, M\'{e}xico}

\author{A. Mart\'{i}n-Ruiz}
\email{alberto.martin@nucleares.unam.mx}
\affiliation{Instituto de Ciencias Nucleares, Universidad Nacional Aut\'{o}noma de M\'{e}xico, 04510 Ciudad de M\'{e}xico, M\'{e}xico}
\affiliation{Centro de Ciencias de la Complejidad, Universidad Nacional Aut\'{o}noma de M\'{e}xico, 04510 Ciudad de M\'{e}xico, M\'{e}xico}

\author{A. Frank}
\email{frank@nucleares.unam.mx}
\affiliation{Instituto de Ciencias Nucleares, Universidad Nacional Aut\'{o}noma de M\'{e}xico, 04510 Ciudad de M\'{e}xico, M\'{e}xico}
\affiliation{Centro de Ciencias de la Complejidad, Universidad Nacional Aut\'{o}noma de M\'{e}xico, 04510 Ciudad de M\'{e}xico, M\'{e}xico}
\affiliation{El Colegio Nacional, Ciudad de M\'{e}xico, M\'{e}xico}

\author{E. Nahmad-Achar}
\email{nahmad@nucleares.unam.mx}
\affiliation{Instituto de Ciencias Nucleares, Universidad Nacional Aut\'{o}noma de M\'{e}xico, 04510 Ciudad de M\'{e}xico, M\'{e}xico}

\begin{abstract}
	A quantum version of the Monty Hall problem is proposed inspired by an experimentally-feasible, quantum-optical set-up that resembles the classical game. The expected payoff of the player is studied by analyzing the classical expectation values of the obtained quantum probabilities. Results are examined by considering both entanglement and non-entanglement between player and host, and using two different approaches: random and strategy-based. We also discuss the influence of noise on the game outcome when the parties play through a noisy quantum channel. The experimental set-up can be used to quickly verify the counter-intuitive result of the Monty Hall problem, adding pedagogic value to the proposal.
\end{abstract}

\maketitle

\section{Introduction}

Inspired by the applications of game theory to the study of classical information problems, quantum information theorists began to include quantum probability amplitudes and quantum entanglement into classical game theory \cite{1,2,3,4,5,6,7}, creating what is now known as quantum game theory. The main motivation behind this new area was, and still is, the applications it has in secure quantum communications, as quantum eavesdropping can be treated as a game in which the spy's goal is to extract the maximum amount of information from a quantum communication channel.

In 1999, Meyer provided the techniques to include quantum strategies into classical game theory \cite{1}, showing that the use of these actually increases the expected payoffs of the players. In the same year, motivated by the lack of an unconditionally secure remote gambling, Goldenberg et al. presented a protocol that allows two remote parties to play a quantum fair-gambling game \cite{2}. Later on, a quantum version of the famous game, the prisoner's dilemma, was discussed by Eisert et al. and Benjamin and Hayden in Refs. \cite{3,4,5}. Collective quantum games, in which more than two players take part, were also studied by Benjamin and Hayden in Ref. \cite{6}, concluding that quantum entanglement enables different kinds of cooperative behavior, preventing players from betraying one another. One interesting problem that arises in quantum game theory is the relation between the classical and quantum results. This issue is briefly discussed by van Enk and Pike in Ref. \cite{7}. In their paper, they analyze to what extent the quantum solution of the prisoner's dilemma solves the classical problem.

Another well-known dilemma in classical information theory is the Monty Hall problem because of its sharply counter-intuitiveness. It has been suggested that a quantum version of the Monty Hall problem may be of interest in the study of quantum strategies of quantum measurement \cite{Chuan}. Also, it has been recently used to reformulate the Pusey-Barret-Rudolph (PBR) theorem \cite{PBR}, as well as to improve the reliability of quantum teleportation \cite{QReports}. PBR addresses the question of whether a quantum state corresponds to a $\psi$-ontic model or to a $\psi$-epistemic model. When expressed as a Monty Hall game, wining probabilities for switching doors depend on whether it is a $\psi$-ontic or a $\psi$-epistemic game \cite{QReports}. Other attempts at a quantum version of the Monty Hall problem can be found in the literature, see for example Refs. \cite{Chuan,Abbott,Ariano,Zander,Gawron,Khan, Kurzyk, Paul}. All these studies show that there is not a unique way to formulate a quantum version of this classical game. In this paper we propose a quantum version of the Monty Hall problem inspired by an experimentally-feasible, quantum-optical set-up that resembles the classical game with the addition of some quantum features. In order to introduce our model, in the following we summarize both the classical and (the various versions of) the quantum Monty Hall game.

\subsubsection*{Classical Monty Hall problem}

The Monty Hall problem is a famous, seemingly paradoxical problem in probability \cite{8,9,10}; closely related to other dilemma problems like the ``three prisoners problem'' and the ``Bertrand's box paradox''. It describes a contest in which a player is asked to choose between three boxes. Inside one of the boxes, a prize was randomly placed beforehand. There are two main characters in this contest: the host (Monty Hall), who knows in which box the prize is, and the player, who does not have any information about its location.

The contest begins with the player choosing (but not opening) one of the boxes. If the chosen box is the one with the prize inside, the host, who knows where the prize hides, randomly opens one of the two empty boxes. On the other hand, if the player chooses one of the empty boxes, the host opens the other remaining empty box. In both cases the host shares this information with the player. Lastly, the host asks the player if she wants to open her initial choice or prefers to open the other box that remains closed. The apparent paradox results from the fact that, when doing the calculations, it is found that the probability of the player finding the prize in the box she initially chose is $\frac{1}{3}$, while the probability of finding the prize if she decides to open the other box is $\frac{2}{3}$.

The above result can be thought of as follows: the location of the prize, with a probability of $\frac{1}{3}$ for each box, and the initial choice of the player, also with a probability of $\frac{1}{3}$ for each box, are independent events; therefore, the probability of the prize being in box $j$ and the player initially choosing box $i$ is $P(i,j) = \frac{1}{9}$ for all $i,j = 1,2,3$. Table \ref{tb:1} shows the elements $(i,j)$ of the corresponding sample space. The cases in which the player initially chooses the box in which the prize is located, that is, the cases in which the player wins if she decides to open her first option, are of the form $(j,j)$. The elements $(i,j)$ with $i\neq j$ represent, after the host has opened box $k$ ($k\neq i,j$), the cases in which the player wins if she decides to change her initial choice and open the other box. The probability $P_{ns}$ of the player winning by not switching her initial choice, is calculated by adding the probabilities of the elements corresponding to that event, that is:
\begin{align}
P_{ns} = P(1,1) + P(2,2) + P(3,3) = \frac{1}{3}. \label{eq:a}
\end{align}
Analogously, the probability $P_{s}$ of the player winning by switching her initial choice, is:
\begin{align}
P_{s} = P(1,2) + P (1,3) + P(2,1) + P(2,3) + P(3,1) + P(3,2) = \frac{2}{3}.\label{eq:b}
\end{align}

\begin{table}[t]
	\begin{tabular}{ccc}
		$(1,1)$\quad  $(1,2)$\quad  $(1,3)$  &  \\ 
		$(2,1)$\quad  $(2,2)$\quad  $(2,3)$  &  \\ 
		$(3,1)$\quad  $(3,2)$\quad  $(3,3)$  & 
	\end{tabular}
	\caption{Elements of the sample space of the Monty Hall problem. The prize being in box $j$ and the player initially choosing box $i$ is represented as $(i,j)$. \label{tb:1}}
\end{table}

The Monty Hall problem can be generalized to include an arbitrary number of boxes, more players (all win the prize if they choose the correct box) and more empty boxes to be opened by the host. In this case, using a similar analysis as before, the probabilities are found to be:

\begin{equation}
P_{ns} = \frac{1}{n},
\end{equation}
\begin{equation}
P_{s} = \left( \frac{n-1}{n-m-1}\right)  \frac{1}{n},
\end{equation}
where $n$ is the total number of boxes and $m$ is the number of empty boxes to be opened by the host. Here, $m$ is restricted to $0\leq m \leq n-k$, where $k$ is the total number of players (counting the host as a player).

\subsubsection*{Quantum Monty Hall game}

In quantum game theory, the construction of a quantum version of a classical game is an entirely subjective task, usually referring to considering the different elements of the classical game (for example, the location of the prize, the opening of a box by the host, etc.) as elements present in the study of a quantum system (such as superpositions and projective measurements, for instance). Consequently, to this day different approaches and quantum versions of the Monty Hall problem have already been proposed \cite{Chuan,Abbott,Ariano,Zander,Gawron,Khan, Kurzyk}, and there is even a quantum algorithm developed so that two persons can play a version of the Monty Hall quantum game on a quantum computer \cite{Paul}. Some of the most interesting quantum versions are those of Flitney and Abbot \cite{Abbott}, D'Ariano et al. \cite{Ariano} and C.-F. Li \cite{Chuan}. Let us briefly summarize these three versions of the quantum Monty Hall game.

\begin{itemize}

\item In Ref. \cite{Abbott}, Flitney and Abbott describe a quantum version in which the location of the prize, the initially chosen box of the player and the box opened by the host, are each represented by a state in a three dimensional Hilbert space: $\mathcal H_{a}$, $\mathcal H_{b}$ and $\mathcal H_{o}$, respectively. The full system is initially prepared in an arbitrary state $\left| \psi \right\rangle \in \mathcal H = \mathcal H_{o} \otimes \mathcal H_{b} \otimes \mathcal H_{a}$, a feature that does not have a classical analogy. After this initial preparation, the game begins: the host hides the prize by acting a unitary operator on $\mathcal H_{a}$, the act of the player choosing a box is also implemented by acting a unitary operator on $\mathcal H_{b}$ and the opening of a box by the host is performed too by acting a unitary operator on the full space $\mathcal H$. Lastly, the decision of switching between boxes is made by acting a superposition of two operators, a switching operator and a not-switching operator, on the full space $\mathcal H$, allowing a non-classical feature: a superposition of switching and not switching between boxes. The authors conclude that if the host has access to quantum strategies and the player does not, the former can make the game fair with a expected payoff of $1/2$ for each player. Otherwise, if the player has access to quantum strategies and the host does not, then the player can win the game all the times.

\item In Ref. \cite{Ariano}, D'Ariano et al. present a scheme in which they represent the location of the prize as a state in a three dimensional space $\mathcal H$, information about the state is given to the host classically or via measurements on an ancillary system entangled with the prize. The initial choice of the player is in this case, not an operation, but a simple choice of a state $\left|p_{i} \right\rangle \in \mathcal H$. The opening of an empty box by the host, is performed as a projective measurement, reducing the space $\cal H$ to a two-dimensional space $\mathcal H _{p}$ spanned by $\left|p_{i} \right\rangle$ and the prize state. Lastly, the player can choose to stay with her initial choice or to change to a different state $\left|p_{f} \right\rangle \in \mathcal{H}_p$. An interesting variation of the Monty Hall game which is closely related with that of D'Ariano et al. is the Ignorant Monty Hall game \cite{QReports}, where  the host does not know where the prize is and can accidentally reveal it. In contrast to the original version where the player expects a payoff of 1/3 if sticking to the initial choice and 2/3 if switching, in this variation the contestant probability of winning is the same whether the contestant chooses to switch the door or not.

\item A different way to implement quantum strategies in the Monty Hall game is proposed by C.-F. Li et al. in Ref. \cite{Chuan}. There are one quantum particle (the prize) and three boxes $ \left| 0 \right> $, $\left| 1 \right>$, and $\left| 2 \right> $. The host puts the particle into the boxes, which may be a superposition state like $\left| \psi \right> _{p} = \frac{1}{\sqrt{3}} ( \left| 0 \right> + \left| 1 \right> + \left| 2 \right> )$. The player now picks a box, for instance $\left| 0 \right>$, and has 1/3 chance to win. If the host reveals no particle in $\left| 2 \right>$, the state of the particle may be described by the density matrix $\rho _{0} = \frac{1}{3} \left| 0 \right> \left< 0 \right| + \frac{2}{3} \left| 1 \right> \left< 1 \right| $. Now the host is allowed to perform a von Neumann measurement on the particle (in orthogonal bases which are linear superpositions of $\left| 0 \right>$ and $\left| 1 \right>$), thus reducing the problem to a coin tossing game. Hence, the conclusion is the same as in the Ignorant Monty Hall game described above: 1/2 of probability of being correct by either staying or switching.

\end{itemize}

In this work we introduce a different approach to the Monty Hall problem which is inspired by a quantum-optical set-up that resembles the classical game. A nonlinear crystal allows us to have entanglement between the prize's location and the initial choice of the player, while the use of polarized beam splitters and polarization rotators enable the superposition feature of both the prize's location and the initial choice of the player. In our quantum version, however, the decision of switching between boxes is binary and based on the classical sample space in table \ref{tb:1}. We will also analyze the  influence of noise on the game outcome.

\section{Proposed Quantum Version} \label{PQS}

The system under study will be modeled by two Hilbert spaces $\mathcal H_{a}$ and $\mathcal H_{b}$, each with a respective orthonormal basis $\left\lbrace \left| 1_{a} \right\rangle,\left| 2_{a} \right\rangle,\left| 3_{a} \right\rangle \right\rbrace$ and $\left\lbrace \left| 1_{b} \right\rangle,\left| 2_{b} \right\rangle,\left| 3_{b} \right\rangle \right\rbrace$. States in $\mathcal H_{a}$ correspond to the initially chosen box by the player (which we will refer to as Alice). Notice that the initial choice of Alice can be a superposition of boxes. On the other hand, states in $\mathcal H_{b}$ correspond to the prize's location, initially prepared by the host (which we will refer to as Bob). Analogously, the prize can also be in a superposition of boxes.

The game begins with Bob preparing the prize in a state

\begin{equation}
\left| \psi_{b} \right\rangle = \sum^{3}_{i=1} b_{i}\left| i_{b} \right\rangle,\label{choice_b}
\end{equation}
which Alice will have no information about. She then proceeds to choose a box, i.e. a state in $\mathcal H_{a}$:

\begin{equation}
\left| \psi_{a} \right\rangle = \sum^{3}_{i=1} a_{i}\left| i_{a} \right\rangle.\label{choice_a}
\end{equation}

The state that describes both the prize and Alice's initial choice is thus 

\begin{equation}
\left| \psi_{0} \right\rangle = \left| \psi_{a} \right\rangle \otimes \left| \psi_{b} \right\rangle \in \mathcal H_{a} \otimes \mathcal H_{b}. \label{eq:1}
\end{equation}

Once Alice has chosen a box, Bob has to do something analogous to open one of them. A useful way (for our proposed experimental set-up) of mathematically doing this, is to apply the door-opening operator

\begin{equation}
\hat{D}_{o} := \cos(\varphi_1) \left| 1_b \right\rangle \left\langle 1_b \right| + \sin(\varphi_2) \left| 2_b \right\rangle \left\langle 2_b \right| + \sin(\varphi_3) \left| 3_b \right\rangle \left\langle 3_b \right|\label{doo}
\end{equation}
to $\left| \psi_{b} \right\rangle$, where $\varphi_{1},\varphi_{2},\varphi_{3} \in \left[ 0, \frac{\pi}{2} \right]$. The door-opening operator must also satisfy a two-doors-remained-closed condition, which can be modeled as

\begin{equation}
\cos^{2}(\varphi_1) + \sin^{2}(\varphi_2) + \sin^{2}(\varphi_3) = 2.\label{tdrcc}
\end{equation}

There are two main things worth noticing here. The first is that the door-opening operator acts as a projection onto the two-dimensional subspace generated by $\left\lbrace \left| 2_{b} \right\rangle,\left| 3_{b} \right\rangle \right\rbrace$, $\left\lbrace \left| 1_{b} \right\rangle,\left| 3_{b} \right\rangle \right\rbrace$ and $\left\lbrace \left| 1_{b} \right\rangle,\left| 2_{b} \right\rangle \right\rbrace$ when ($\varphi_{1} = \frac{\pi}{2},\varphi_{2} = \frac{\pi}{2},\varphi_{3} = \frac{\pi}{2}$), ($\varphi_{1} = 0,\varphi_{2} = 0,\varphi_{3} = \frac{\pi}{2}$) and ($\varphi_{1} = 0,\varphi_{2} = \frac{\pi}{2},\varphi_{3} = 0$) respectively. The second is that the application of $\hat{D}_{o}$ on $\left| \psi_{b} \right\rangle$ de-normalizes it. Thus, in order to maintain the interpretation of the inner product as a probability amplitude, the resulting state must be renormalized, leading to define

\begin{equation}
\sum^{3}_{i=1} \beta_{i}\left| i_{b} \right\rangle = \frac{\hat{D}_{o} \left| \psi_{b} \right\rangle}{\sqrt{\left\langle \psi_b \right| \hat{D}^{\dagger}_{o} \hat{D}_{o} \left| \psi_{b} \right\rangle}},\label{betas}
\end{equation}
for some $\beta_{i}\in \mathbb{C}$ such that $\left| \beta_{1} \right| ^{2} + \left| \beta_{2} \right| ^{2} + \left| \beta_{3} \right| ^{2} = 1$. The composite state of Alice's initial choice and the prize's location thus becomes

\begin{equation}
\left| \psi \right\rangle = \sum^{3}_{i=1} \sum^{3}_{j=1} a_{i} \beta_{j} \left| i_{a}, \; j_{b} \right\rangle. \label{obs}
\end{equation}

In analogy with the classical case, and as it was shown in table \ref{tb:1}, the states $\left\lbrace \left| 1_{a} \; 1_{b} \right\rangle,\left| 2_{a} \; 2_{b} \right\rangle,\left| 3_{a} \; 3_{b} \right\rangle \right\rbrace$ correspond to Alice winning with her initial choice, while the states $\left\lbrace \left| 1_{a} \; 2_{b} \right\rangle,\left| 1_{a} \; 3_{b} \right\rangle,\left| 2_{a} \; 1_{b} \right\rangle,\left| 2_{a} \; 3_{b} \right\rangle,\left| 3_{a} \; 1_{b} \right\rangle,\left| 3_{a} \; 2_{b} \right\rangle \right\rbrace$ correspond to Alice winning by switching her initial choice. Therefore, the probability of Alice winning by not switching and the probability of her winning by switching are respectively

\begin{equation}
P_{ns} = \sum^{3}_{i=1} \left| a_{i} \beta_{i} \right|^{2}, \label{eq:2}
\end{equation}

\begin{equation}
P_{s} = \sum^{3}_{i,j=1 \; (i\neq j)} \left| a_{i} \beta_{j} \right|^{2}. \label{eq:3}
\end{equation}

The classical Monty Hall problem may be thought of using this quantum version by restricting Alice and Bob to only use ``classical'' states $\left\lbrace \left| 1_{a} \right\rangle,\left| 2_{a} \right\rangle,\left| 3_{a} \right\rangle \right\rbrace$ and $\left\lbrace \left| 1_{b} \right\rangle,\left| 2_{b} \right\rangle,\left| 3_{b} \right\rangle \right\rbrace$ respectively. Nevertheless, expressions \eqref{eq:2} and \eqref{eq:3} would only yield the objective quantum probability (i.e. $1$ or $0$) of Alice winning given her choices and the state of the prize. In order to calculate the subjective classical probability (i.e. the one due to the lack of information from Alice) and obtain the same results as in \eqref{eq:a} and \eqref{eq:b}, one must perform a classical probability calculation (using table \ref{tb:1} for example).

There is, however, a selection of quantum states by Alice and Bob that actually resembles the classical problem, a semi-classical case: First, Bob prepares the prize in the state with $b_{i}=\frac{1}{\sqrt{3}}$ (i.e. equally distributed over the three boxes), then Alice chooses the state with $a_{i}=\frac{1}{\sqrt{3}}$ (i.e. the same confidence in each box) and finally, Bob applies the door-opening operator $\hat{D}_{o}$ with ($\varphi_{1} = \frac{\pi}{2},\varphi_{2} = \frac{\pi}{2},\varphi_{3} = \frac{\pi}{2}$) or ($\varphi_{1} = 0,\varphi_{2} = 0,\varphi_{3} = \frac{\pi}{2}$) or ($\varphi_{1} = 0 , \varphi_{2} = \frac{\pi}{2},\varphi_{3} = 0$). In this case, equations \eqref{eq:2} and \eqref{eq:3} give respectively $\frac{1}{3}$ and $\frac{2}{3}$.

The semi-classical case discussed above points out to a classical interpretation of the amplitudes $a_{i}$ and $b_{i}$ in our scheme, being $\left| a_{i} \right|^2$ the probability of Alice initially choosing box $i$ and $\left| b_{i} \right|^2$ the probability of the prize being placed in box $i$. This means that our quantum version, until now, can be entirely reproduced in classical probability theory by the classical Monty Hall problem or by analyzing a non-symmetric case: the host is more inclined to hide the prize in certain box and the player has some kind of bias towards initially choosing one of the boxes. However, it is worth mentioning that expression \eqref{obs} assumes $\left| \psi_{0} \right\rangle$ is a separable state in $\mathcal H_{a} \otimes \mathcal H_{b}$ (equation \eqref{eq:1}). This is a fully classical restriction, one we do not have to follow in the quantum realm. For a general state

\begin{equation}
\left| \psi \right\rangle = \sum_{i,j=1}^{3} \gamma_{ij} \left| i_{a}, \; j_{b} \right\rangle \in \mathcal H_{a} \otimes \mathcal H_{b}, \label{eq:un}
\end{equation}
the probability of Alice winning by not switching and the probability of her winning by switching are respectively

\begin{equation}
P_{e,ns} = \sum^{3}_{i=1} \left| \gamma_{ii}\right|^{2}, \label{eq:ens}
\end{equation}

\begin{equation}
P_{e,s} = \sum^{3}_{i,j=1 \; (i\neq j)} \left| \gamma_{ij}\right|^{2}. \label{eq:es}
\end{equation}

The consideration of entanglement in equation \eqref{eq:un} between the prize's location and the player's initial choice would make the game unfair. It is here considered for the purpose of extending our framework to construct a secure communication protocol, as will be discussed below.

In the next section we describe the quantum-optical set-up in which our quantum version is inspired, and use it to analyze both a separable and an entangled initial state $\left| \psi_{0} \right\rangle$.

\section{Experimental Realization}

\begin{figure*}[t]
	\begin{centering}
		\includegraphics[scale=1]{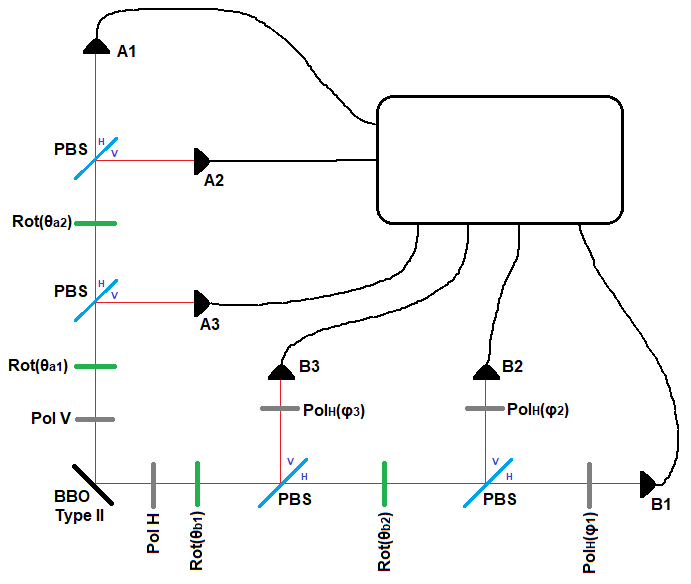}
		\par\end{centering}
	
	\caption{\label{fig:1} Diagram of the experimental set-up proposed for the Monty Hall problem. The BBO type II crystal produces a pair of entangled photons in both position and polarization. The vertical arm of the set-up corresponds to Alice's subsystem, while the horizontal one corresponds to Bob's. ``Pol V'' and ``Pol H'' represent vertical and horizontal polarizers respectively, ``Rot($\theta$)'' represents a polarization rotator of an angle $\theta$, ``PBS'' represents a polarized beam splitter, ``Pol$_{H}(\varphi)$'' represents a polarizer at an angle $\varphi$ respect to the horizontal polarization. Detectors A1, A2 and A3 correspond to Alice's initial choice of a box, while the detectors B1, B2 and B3 correspond to the doors in which the prize is prepared.}
\end{figure*}

Figure \ref{fig:1} shows a diagram of a quantum-optical approach to an experimental realization of our proposed quantum version for the Monty Hall problem. In it, detectors A1, A2 and A3 represent Alice's initial choice of a box, while detectors B1, B2 and B3 represent the boxes where the prize is hidden.

The game starts with a pair of polarization-entangled photons produced via spontaneous parametric down-conversion in a Beta Barium Borate (BBO) type II crystal. Photon A, represented by the vertical output of the BBO crystal in the diagram, corresponds to the system available to Alice and modeled in $\mathcal H_{a}$. While photon B, represented by the horizontal output of the BBO crystal in the diagram, corresponds to the system available to Bob and modeled in $\mathcal H_{b}$. The state of the composite system at this stage is given by the entangled state

\begin{equation}
\left| \phi_0 \right\rangle = \frac{1}{\sqrt{2}} \left( \left| V_{1}, \; H_{1} \right\rangle + \left| H_{1}, \; V_{1} \right\rangle \right)  \in \mathcal H_{a} \otimes \mathcal H_{b}, \label{entan}
\end{equation}
where $V$ and $H$ stand for vertical and horizontal polarization respectively, and the subindex $1$ represents the label of the detector to which the photon is heading (A1 for photon A and B1 for photon B).

Firstly, a vertical polarizer (Pol V) and an horizontal polarizer (Pol H) are respectively placed in photon A's and photon B's path. These polarizers allow to control entanglement between the two subsystems. If the polarizers are present, the entanglement between photons A and B is lost, as the state $\left| \phi_0 \right\rangle$ reduces to $\left| V_{1}, \; H_{1} \right\rangle$. On the other hand, if the polarizers are removed, then the system remains entangled in the state $\left| \phi_0 \right\rangle$.

Both entangled and non-entangled cases are analyzed, but for the sake of simplicity, let us describe just the case where the polarizers are placed and the initial state of the system is decribed by

\begin{equation}
\left| \phi_0 \right\rangle = \left| V_{1}, \; H_{1} \right\rangle. \label{nen}
\end{equation}
In order to obtain the expressions corresponding to the entangled case, the operations we will be describing must also be applied to the term $\left| H_{1}, \; V_{1} \right\rangle$.

The next device placed in the photons path is a polarization rotator (Rot($\theta$)) with rotation angles $\theta_{a1}$ and $\theta_{b1}$ for photon A and photon B respectively. By choosing an angle $\theta_{b1}$, Bob is fixing the probability amplitude $b_3$ in \eqref{choice_b}. Analogously, by choosing an angle $\theta_{a1}$, Alice is fixing the probability amplitude $a_3$ in eq. \eqref{choice_a}.

The polarization-rotator linear operator performs the operations

\begin{align}
Rot(\theta)\left| V \right\rangle & = \cos{\theta}\left| V \right\rangle - \sin{\theta}\left| H \right\rangle \label{eq:4} \\ Rot(\theta)\left| H \right\rangle & = \cos{\theta}\left| H \right\rangle + \sin{\theta}\left| V \right\rangle,\label{eq:5}
\end{align}
changing the initial state $\left| \phi_0 \right\rangle$ to

\begin{equation}
\left| \phi_1 \right\rangle = \left( \cos{\theta_{a1}} \left| V_{1}\right\rangle - \sin{\theta_{a1}} \left| H_{1}\right\rangle \right) \otimes \left( \cos{\theta_{b1}} \left| H_{1}\right\rangle + \sin{\theta_{b1}} \left| V_{1}\right\rangle \right).
\end{equation}

Next, each of the photons encounter a polarized beam splitter (PBS), positioned to reflect the vertical component of the polarization and transmit the horizontal one. These first PBS's reflect the vertical component of the polarization of photons A and B towards detectors A3 and B3 respectively, performing the operation

\begin{equation}
PBS \left( \alpha \left| V_1 \right\rangle + \beta \left| H_1 \right\rangle \right)  = \alpha \left| V_3 \right\rangle + \beta \left| H_1 \right\rangle.
\end{equation}
The state of the system after these first PBS's is then

\begin{equation}
\left| \phi_2 \right\rangle = \left( \cos{\theta_{a1}} \left| V_{3}\right\rangle - \sin{\theta_{a1}} \left| H_{1}\right\rangle \right) \\ \otimes \left( \cos{\theta_{b1}} \left| H_{1}\right\rangle + \sin{\theta_{b1}} \left| V_{3}\right\rangle \right).\label{eq:6}
\end{equation}

It is worth mentioning that, strictly speaking, the operation performed by the PBS requires an ancillary input in order to have two outputs, accounting for the four faces of a BS. Since in this work we are just using a PBS as a controlled gate and the proposed experimental set-up does not have any interferometer-like behavior, this more-formal mathematical treatment of a PBS is not necessary and will not be used.

In order for Alice and Bob to fix amplitudes $a_1$, $a_2$ in \eqref{choice_a} and $b_1$, $b_2$ in \eqref{choice_b}, another polarization rotators with angles $\theta_{a2}$ and $\theta_{b2}$ are respectively placed in the path of the horizontal component of photons A and B. Applying the polarization-rotator operator (equations \eqref{eq:4} and \eqref{eq:5}) to state $\left| \phi_2 \right\rangle$ with angles $\theta_{a2}$ and $\theta_{b2}$, the state of the system becomes

\begin{multline}
\left| \phi_3 \right\rangle = ( \cos{\theta_{a1}} \left| V_{3}\right\rangle - \sin{\theta_{a1}} \sin{\theta_{a2}} \left| V_{1}\right\rangle - \sin{\theta_{a1}}\cos{\theta_{a2}} \left| H_{1}\right\rangle ) \\ \otimes ( \sin{\theta_{b1}} \left| V_{3} \right\rangle + \cos{\theta_{b1}} \sin{\theta_{b2}} \left| V_{1}\right\rangle + \cos{\theta_{b1}}\cos{\theta_{b2}} \left| H_{1}\right\rangle ).
\end{multline}

Then, a polarized beam splitter (PBS) is placed in both photons paths. These last PBS's are positioned to reflect the vertical component of the polarization towards detectors A2 and B2. So the state of the system becomes

\begin{multline}
\left| \phi_4 \right\rangle = ( \cos{\theta_{a1}} \left| V_{3}\right\rangle - \sin{\theta_{a1}} \sin{\theta_{a2}} \left| V_{2}\right\rangle - \sin{\theta_{a1}}\cos{\theta_{a2}} \left| H_{1}\right\rangle) \\ \otimes ( \sin{\theta_{b1}} \left| V_{3} \right\rangle + \cos{\theta_{b1}} \sin{\theta_{b2}} \left| V_{2}\right\rangle + \cos{\theta_{b1}}\cos{\theta_{b2}} \left| H_{1}\right\rangle ). \label{eq:7}
\end{multline}

Notice equation \eqref{eq:1} is analogous to equation \eqref{eq:7} with

\begin{align}
a_{1} & = - \sin{\theta_{a1}}\cos{\theta_{a2}},\label{a1}\\
a_{2} & = - \sin{\theta_{a1}} \sin{\theta_{a2}},\label{a2}\\
a_{3} & = \cos{\theta_{a1}},\label{a3}\\
b_{1} & = \cos{\theta_{b1}} \cos{\theta_{b2}},\\
b_{2} & = \cos{\theta_{b1}} \sin{\theta_{b2}},\\
b_{3} & = \sin{\theta_{b1}}.
\end{align}

The process described so far has been to prepare the state of the system as in expression \eqref{eq:1}. The next step in the Monty Hall problem is for Bob to open one box. In order to model this door-opening procedure in our experimental set-up, three polarizers (Pol$_{H}(\varphi_{i})$) at angles $\varphi_ {i}$ respect to the horizontal polarization, are placed before the detectors B$_i$. Strictly speaking, a polarizer performs a dissipative operation, since it absorbs part of the radiation that arrives to it. However, we are only interested in photons that account for coincidences between detectors A and B, i.e. photons that do reach detectors B1, B2 or B3. This is because these photons are the ones that allow us to measure the probabilities in an analogous way as we did classically with table \ref{tb:1}. Therefore, we can include the effect of the polarizers by changing the probability amplitudes $b_ {i}$ in Bob's subsystem as

\begin{align}
b^{\prime}_{1} = b_{1}\cos{\varphi_{1}},\\
b^{\prime}_{2} = b_{2}\sin{\varphi_{2}},\\
b^{\prime}_{3} = b_{3}\sin{\varphi_{3}},
\end{align}
which is equivalent to applying the door-opening operator \eqref{doo} to Bob's subsystem, using the two-doors-remained-closed condition expressed in \eqref{tdrcc} as we only need to block one third of the photons.

Then, as it was described in the previous section, we must renormalize the state, which is experimentally justified by the fact that we are calculating the probability of measuring the state $\left| i_{a}, \; j_{b} \right\rangle$ as

\begin{equation}
P_{i,j} = \frac{C_{i,j}}{\displaystyle\sum_{m,n=1}^{3}C_{m,n}},
\end{equation}
where $C_{m,n}$ stands for the number of coincidences between detectors A$m$ and B$n$.

Renormalization of the state leads to the new probability amplitudes $\beta_i$ (defined in expression \eqref{betas}) for Bob's subsystem:

\begin{equation}
\beta_{i} = \frac{b^{\prime}_{i}}{\sqrt{(b^{\prime}_{1})^{2}+(b^{\prime}_{2})^{2}+(b^{\prime}_{3})^{2}}}.
\end{equation}

In the case when the first polarizers (Pol V and Pol H) are removed, and the initial state $\left| \phi_0 \right\rangle$ is as in equation \eqref{entan}, the non-normalized amplitudes with respect to the basis $\left\lbrace \left| i_{a}, \; j_{b}\right\rangle \right\rbrace^{i=1,2,3}_{j=1,2,3}$, after applying the door-opening operator to the resulting state, turn out to be

\begin{align}
c_{11} & = \frac{-1}{\sqrt{2}} \cos{\theta_{a2}} \cos{\theta_{b2}} \sin{(\theta_{a1}+\theta_{b1})} \cos{\varphi_{1}}, \\
c_{12} & = \frac{-1}{\sqrt{2}} \cos{\theta_{a2}} \sin{\theta_{b2}} \sin{(\theta_{a1}+\theta_{b1})} \sin{\varphi_{2}}, \\
c_{13} & = \frac{1}{\sqrt{2}} \cos{\theta_{a2}} \cos{(\theta_{a1}+\theta_{b1})} \sin{\varphi_{3}}, \\
c_{21} & = \frac{-1}{\sqrt{2}} \sin{\theta_{a2}} \cos{\theta_{b2}} \sin{(\theta_{a1}+\theta_{b1})} \cos{\varphi_{1}},\\
c_{22} & = \frac{-1}{\sqrt{2}} \sin{\theta_{a2}} \sin{\theta_{b2}} \sin{(\theta_{a1}+\theta_{b1})} \sin{\varphi_{2}},\\
c_{23} & = \frac{1}{\sqrt{2}} \sin{\theta_{a2}} \cos{(\theta_{a1}+\theta_{b1})} \sin{\varphi_{3}},\\
c_{31} & = \frac{1}{\sqrt{2}} \cos{\theta_{b2}} \cos{(\theta_{a1}+\theta_{b1})} \cos{\varphi_{1}},\\
c_{32} & = \frac{1}{\sqrt{2}} \sin{\theta_{b2}} \cos{(\theta_{a1}+\theta_{b1})} \sin{\varphi_{2}},\\
c_{33} & = \frac{1}{\sqrt{2}} \sin{(\theta_{a1}+\theta_{b1})} \sin{\varphi_{3}}.
\end{align}
Thus, the amplitudes $\gamma_{ij}$ necessary to calculate $P_{e,ns}$ and $P_{e,s}$ as in equations \eqref{eq:ens} and \eqref{eq:es} respectively, are

\begin{equation}
\gamma_{ij} = \frac{c_{ij}}{\sqrt{\displaystyle\sum_{i,j=1}^{3} \left|c_{ij}\right|^2}}.
\end{equation}

\section{Noise Effects in the Experimetal Realization}

Noise effects in quantum games has been extensively studied \cite{Chen, Johnson, Flitney, Ozdemir}. In Ref. \cite{Gawron}, Gawron et al. considered the influence of a spontaneous emission channel and a generalized Pauli channel on the quantum Monty Hall Game (within the Flittney and Abbott scheme). In this section, we will discuss the effects of a Pauli channel in our quantum version of the Monty Hall problem.

If Alice and Bob played near each other (in the same lab) the quantum Monty Hall game, using the experimental set-up here proposed, the main source of noise in the obtained results would be the experimental uncertainty associated with each device, which can be addressed using experimental and uncertainty-propagation methods. However, a more interesting source of noise arise if we consider that Alice and Bob wish to play remotely; we discuss this case below.

Let us suppose that the whole vertical arm of the experimental set-up, after the initial vertical polarizer (Pol V), is far away in Alice's lab. The rest of the set-up, along with the computer that counts the coincidences between detectors, stay in Bob's lab. The coincidences-count software would have to be tuned in order to account for the difference in separation between detectors A and B from the computer. In this case, in order for Alice and Bob to play the quantum Monty Hall game, the photon in which Alice applies her operations must travel through a quantum channel connecting the set-ups in both labs, leading to a possible loss of information on the state created by the BBO type II crystal, and affecting the game results. This loss of information is modeled using the Pauli noise:

\begin{equation}
\hat{\rho}_{o} = (1-p_{x}-p_{y}-p_{z})\hat{\rho} + p_{x} \, \hat{\sigma}_{x}\hat{\rho}\hat{\sigma}_{x} + p_{y} \, \hat{\sigma}_{y}\hat{\rho}\hat{\sigma}_{y} + p_{z} \, \hat{\sigma}_{z}\hat{\rho}\hat{\sigma}_{z},
\end{equation}
where $\hat{\rho}_{o}$ is the state at the channel's output, $\hat{\rho}$ is the state at the channel's input, $\sigma_{i}$ are the Pauli matrices, and $p_{x}$, $p_{y}$ and $p_{z}$ are real parameters, between $0$ and $1$ and such that $p_{x}+p_{y}+p_{z} = 1$, related to the fidelity of the quantum channel along the respective axis.

In this work, for simplicity and in order to avoid specifying a fixed configuration of the channel, we consider $p_{x} = p_{y} = p_{z} = \frac{p}{3}$. Furthermore, since just Alice's photon is sent through the quantum channel, it is the only one subject to the noise, thus, in our proposed experimental set-up, the state at the output of the channel is given by 

\begin{equation}
\hat{\rho}_{o} = (1-p)\hat{\rho} + \frac{p}{3}\left[ \left( \hat{\sigma}_{x} \otimes \hat{I} \right) \hat{\rho} \left( \hat{\sigma}_{x} \otimes \hat{I} \right) + \left( \hat{\sigma}_{y} \otimes \hat{I} \right) \hat{\rho} \left( \hat{\sigma}_{y} \otimes \hat{I} \right) + \left( \hat{\sigma}_{z} \otimes \hat{I} \right) \hat{\rho} \left( \hat{\sigma}_{z} \otimes \hat{I} \right) \right], \label{output}
\end{equation}
where $\hat{I}$ is the identity operator.

In our proposed experimental set-up, the possible states of the system at the channel's input are given by

\begin{equation}
\hat{\rho} = \left| V_{1}, \, H_{1} \right\rangle \left\langle V_{1}, \, H_{1} \right|,
\end{equation}
\begin{equation}
\hat{\rho}_{e} = \frac{1}{2} \left( \left| V_{1}, \, H_{1} \right\rangle \left\langle V_{1}, \, H_{1} \right| + \left| V_{1}, \, H_{1} \right\rangle \left\langle H_{1}, \, V_{1} \right| + \left| H_{1}, \, V_{1} \right\rangle \left\langle V_{1}, \, H_{1} \right| + \left| H_{1}, \, V_{1} \right\rangle \left\langle H_{1}, \, V_{1} \right| \right),
\end{equation}
where $\hat{\rho}$ and $\hat{\rho}_{e}$ stand respectively for the non-entangled and entangled cases.

Using equation \eqref{output}, the possible states at the channel's output, for the non-entangled and entangled cases, are respectively

\begin{equation}
\hat{\rho}_{o} = \left(1 - \frac{2p}{3} \right)  \left| V_{1}, \, H_{1} \right\rangle \left\langle V_{1}, \, H_{1} \right| +  \frac{2p}{3} \left| H_{1}, \, H_{1} \right\rangle \left\langle H_{1}, \, H_{1} \right|,
\end{equation}
\begin{multline}
\hat{\rho}_{e,o} = \frac{1}{2} \left[ \left(1 - \frac{2p}{3} \right) \left( \left| H_{1}, \, V_{1} \right\rangle \left\langle H_{1}, \, V_{1} \right| + \left| V_{1}, \, H_{1} \right\rangle \left\langle V_{1}, \, H_{1} \right|\right) \right. \\  \left. + \left(1 - \frac{4p}{3} \right) \left(  \left| V_{1}, \, H_{1} \right\rangle \left\langle H_{1}, \, V_{1} \right| + \left| H_{1}, \, V_{1} \right\rangle \left\langle V_{1}, \, H_{1} \right| \right) +  \frac{2p}{3} \left(  \left| H_{1}, \, H_{1} \right\rangle \left\langle H_{1}, \, H_{1} \right| + \left| V_{1}, \, V_{1} \right\rangle \left\langle V_{1}, \, V_{1} \right| \right)  \right].
\end{multline}

If we denote as $\hat{E}$ the operator that represents all the operations performed by both Alice and Bob in the experimental set-up of the game, the final states of the game for the non-entangled and entangled cases, are respectively given by

\begin{equation}
\hat{\rho}_{f} = \hat{E} \hat{\rho}_{o} \hat{E}^{\dagger},
\end{equation}
\begin{equation}
\hat{\rho}_{e,f} = \hat{E} \hat{\rho}_{e,o} \hat{E}^{\dagger},
\end{equation}
and thus, the probabilities of winning by not switching and by switching for the non-entangled case (expressions \eqref{eq:2} and \eqref{eq:3}), and for the entangled case (expressions \eqref{eq:ens} and \eqref{eq:es}), are generalized as
\begin{align}
    & P_{ns} = \displaystyle\sum_{i=1}^{3} \left\langle i_{a}, \; i_{b} \right| \hat{\rho}_{f} \left| i_{a}, \; i_{b} \right\rangle, \\ & P_{s} = \displaystyle\sum^{3}_{i,j=1 \; (i\neq j)} \left\langle i_{a}, \; j_{b} \right| \hat{\rho}_{f} \left| i_{a}, \; j_{b} \right\rangle, \\ & P_{e,ns} = \displaystyle\sum_{i=1}^{3} \left\langle i_{a}, \; i_{b} \right| \hat{\rho}_{e,f} \left| i_{a}, \; i_{b} \right\rangle, \\ & P_{e,s} = \displaystyle\sum^{3}_{i,j=1 \; (i\neq j)} \left\langle i_{a}, \; j_{b} \right| \hat{\rho}_{e,f} \left| i_{a}, \; j_{b} \right\rangle, 
\end{align}
where, as before, the state $\left| i_{a}, \; j_{b} \right\rangle$ represents the event of Alice detecting a photon in detector Ai and Bob detecting a photon in detector Bj.

\begin{figure*}[!t]
	\begin{centering}
		\includegraphics[scale=0.4]{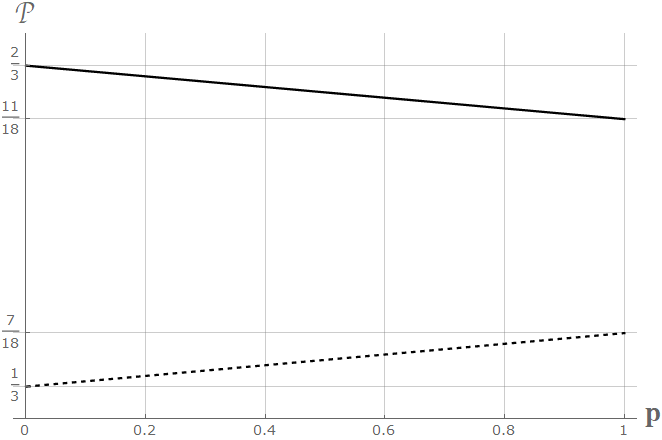}
		\par\end{centering}

	\caption{\label{Pclass_noise} Probabilities of Alice winning the semi-classical Monty Hall game by switching (continuous line) and by not switching (dashed line), as a function of the parameter $p$ related to the fidelity of the quantum channel.}
\end{figure*}

As an example, consider the semi-classical case described in Section \ref{PQS}. Figure \ref{Pclass_noise} shows the probabilities of winning by switching and by not switching as a function of the parameter $p$ related to the fidelity of the quantum channel. Notice that as $p$ increases, the difference between the switching and not switching probabilities decreases, making the decision of switching slightly less meaningful for Alice, from $2$ times better to approximately $1.57$ times better.

\section{Results}

In this section we analyze the expected payoff of Alice (the player) from a frequentist perspective. Namely, if the Monty Hall experimental set-up is played multiple times, does Alice have a better chance of winning the game by not switching (bet for a coincidence between detectors A1/B1, A2/B2 and A3/B3) or by switching (bet for a coincidence between detectors A1/B2, A1/B3, A2/B1, A2/B3, A3/B1 and A3/B2)? We answer this question using two different approaches: random and strategy-based.

\begin{figure*}[!t]
	\begin{centering}
		\includegraphics[scale=0.4]{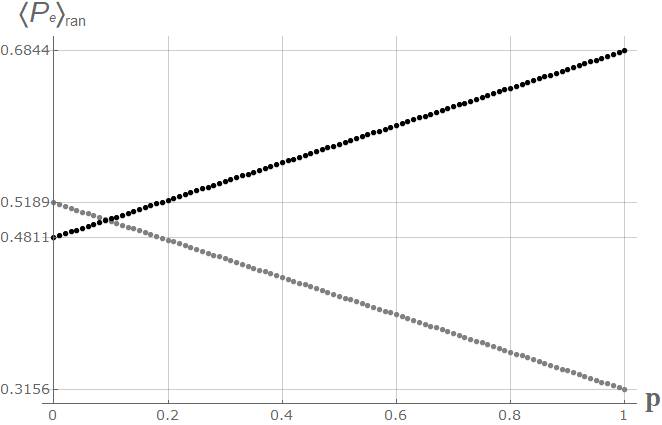}
		\par\end{centering}
	
	\caption{\label{PeRan_noise} Expectation values of the probabilities (random case) of Alice winning by switching (black) and by not switching (gray), as a function of the parameter $p$ related to the fidelity of the quantum channel.}
\end{figure*}

\subsubsection*{Random game}

In the random approach, the parameters of the experiment are considered as random variables with a constant joint probability density function $\rho$. Angles $\theta_{a1}$, $\theta_{a2}$, $\theta_{b1}$ and $\theta_{b2}$ are considered as independent random variables, while the angles associated with the door-opening operator, $\varphi_{1}$, $\varphi_{2}$ and $\varphi_{3}$, are considered as random variables subject to the two-doors-remained-closed condition \eqref{tdrcc}, which restricts its values to the region

\begin{align}
0<\cos{\varphi_{1}}<1,\\
\sin{\varphi_{1}}<\sin{\varphi_{2}}<1,\\
\sin{\varphi_{3}} = \sqrt{\sin^{2}{\varphi_{1}} + \cos^{2}{\varphi_{2}}} \; \; ,
\end{align}
defining the joint probability density function $\rho$ through

\begin{equation}
\frac{1}{\rho} = \int_{0}^{\frac{\pi}{2}} d\theta_{a1} \int_{0}^{\frac{\pi}{2}} d\theta_{a2} \int_{0}^{\frac{\pi}{2}} d\theta_{b1} \int_{0}^{\frac{\pi}{2}} d\theta_{b2} \int_{0}^{\frac{\pi}{2}} d\varphi_{1} \int_{\varphi_{1}}^{\frac{\pi}{2}} d\varphi_{2},
\end{equation}
which leads to
\begin{equation}
\rho = \frac{128}{\pi^6}.
\end{equation}

The probabilities in equations \eqref{eq:2}, \eqref{eq:3}, \eqref{eq:ens} and \eqref{eq:es}, are then functions of these random variables. We use their expectation values with respect to the probability density function $\rho$, as the expected payoff of Alice in each case.

When entanglement is not considered, as in the state described in eq. \eqref{nen}, the expectation values of the probability of winning by not switching and by switching, are respectively

\begin{align}
\left\langle P_{ns} \right\rangle_{ran} & \approx 0.3664, \label{rns} \\
\left\langle P_{s} \right\rangle_{ran} & \approx 0.6336. \label{rs}
\end{align}

Analogously, when entanglement is considered, as in the state described in eq. \eqref{entan}, the expectation values of the probability of winning by not switching and by switching, are respectively

\begin{align}
\left\langle P_{e,ns} \right\rangle_{ran} & \approx 0.5189, \label{rens} \\
\left\langle P_{e,s} \right\rangle_{ran} & \approx 0.4811. \label{res}
\end{align}

When noise is considered, the results obtained without entanglement (expressions \eqref{rns} and \eqref{rs}) remain the same, a somewhat expected result, since we only take the average of all possible parameters without any particular quantum correlation in the system. Nevertheless, the results obtained with entanglement (expressions \eqref{rens} and \eqref{res}) do change as functions of the parameter $p$ related to the fidelity of the channel. Figure \ref{PeRan_noise} shows the effect of noise in the average results of the game when entanglement is considered.

\begin{figure*}[!t]
	\begin{centering}
		\includegraphics[scale=0.4]{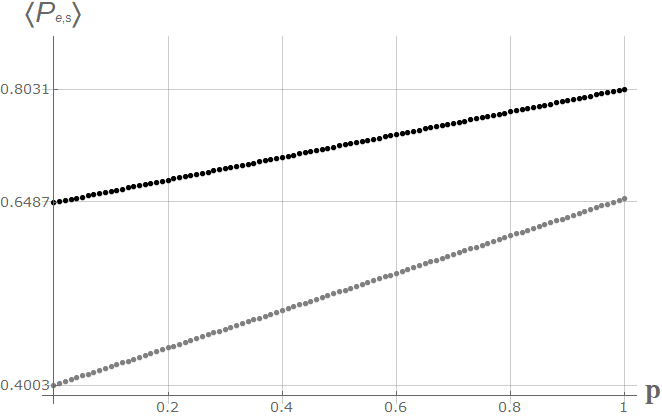}
		\par\end{centering}
	
	\caption{\label{Pe_noise} Maximum (black) and minimum (gray) of the expectation value of the probability of Alice winning by switching when entanglement is considered, as a function of the parameter $p$ related to the fidelity of the quantum channel.}
\end{figure*}

\subsubsection*{Strategy-based game}

In the strategy-based approach, the angles $\theta_{a1}$, $\theta_{a2}$, $\theta_{b1}$ and $\theta_{b2}$ are also considered as independent (in analogy with the classical game) random variables. However, the door-opening parameters, $\varphi_{1}$, $\varphi_{2}$ and $\varphi_{3}$, are left free to be tuned in favor of a certain strategy from Bob (the host). In this approach, the joint probability density function is just

\begin{equation}
\varrho = \frac{16}{\pi^4},
\end{equation}
making the expectation values of the probabilities in equations \eqref{eq:2}, \eqref{eq:3}, \eqref{eq:ens} and \eqref{eq:es}, to be functions of the door-opening parameters $\varphi_{1}$ and $\varphi_{2}$.

\begin{figure*}[!t]
	\begin{centering}
		\includegraphics[scale=0.34]{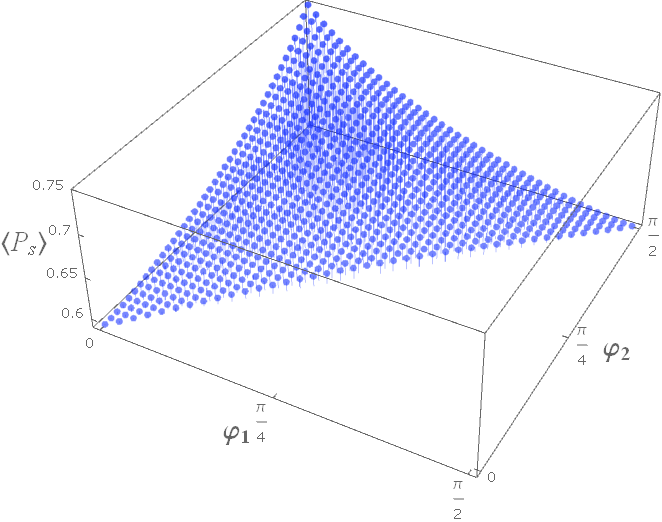} \qquad \quad \includegraphics[scale=0.34]{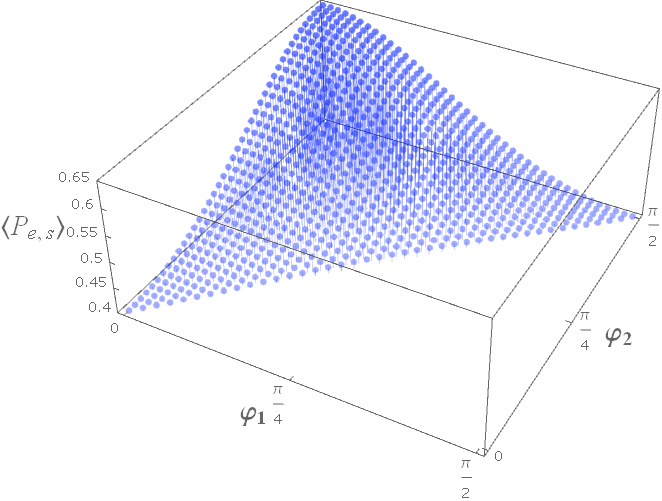}
		\par\end{centering}
	
	\caption{\label{fig:2} Expectation value of the probability of Alice winning by switching, as a function of the door-opening parameters $\varphi_1$ and $\varphi_2$. \textbf{Left:} Non-entangled-case. \textbf{Right:} Entangled case.}
\end{figure*}

Figure \ref{fig:2} shows the graphics of the expectation value of the probability of Alice winning by switching as function of $\varphi_{1}$ and $\varphi_{2}$, for both the entangled $\left\langle P_{e,s}\right\rangle $ and non-entangled cases $\left\langle P_{s}\right\rangle $.

Table \ref{tb:2} shows the approximate maximum and minimum values of $\left\langle P_{s}\right\rangle$, $\left\langle P_{ns}\right\rangle$, $\left\langle P_{e,s}\right\rangle $ and $\left\langle P_{e,ns}\right\rangle $ as well as the approximate corresponding values of $\varphi_{1}$ and $\varphi_{2}$ where that maximum or minimum is reached.

%

\begin{figure*}[!t]
	\begin{centering}
		\includegraphics[scale=0.35]{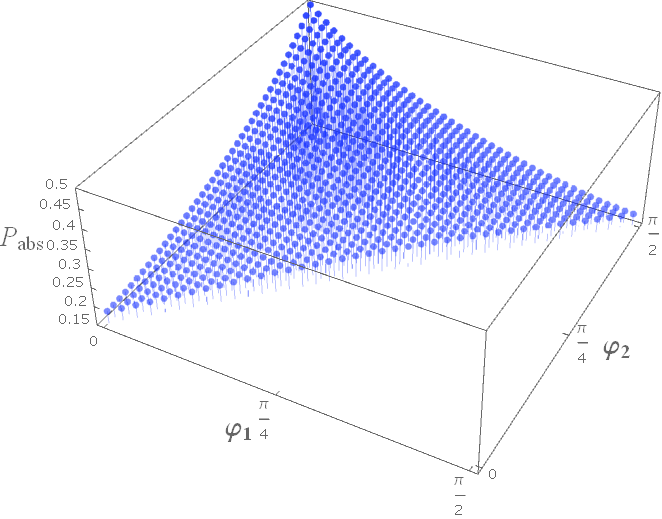} \qquad \quad \includegraphics[scale=0.35]{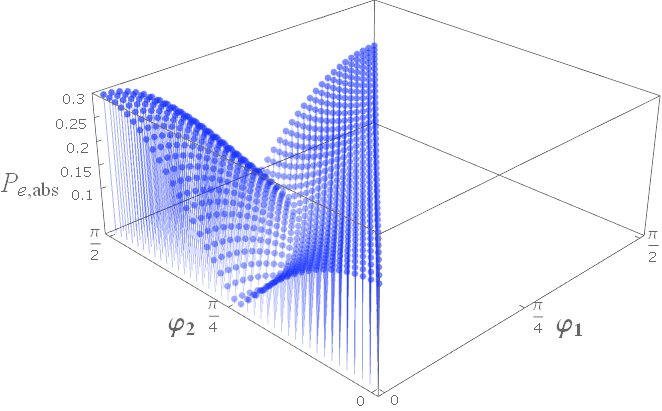}
		\par\end{centering}
	
	\caption{\label{fig:4} Absolute value of the difference between expectation values of the probabilities of Alice winning by not switching and by switching, as a function of the door-opening parameters $\varphi_1$ and $\varphi_2$. \textbf{Left:} Non-entangled-case. \textbf{Right:} Entangled case.}
\end{figure*}


\begin{table}[!h]
	\resizebox{0.3\textwidth}{!}{%
		\begin{tabular}{| c || c | c | c || c | c | c |} 
			\hline
			& $\varphi_{1}$ & $\varphi_{2}$ & min & $\varphi_{1}$ & $\varphi_{2}$ & max \\ 
			\hline \hline
			$\left\langle P_{s}\right\rangle$ & $\frac{\pi}{2}$ & $\frac{\pi}{2}$ & $0.5908$ & $0$ & $\frac{\pi}{2}$ & $0.75$ \\ 
			\hline
			$\left\langle P_{ns}\right\rangle$ & $0$ & $\frac{\pi}{2}$ & $0.25$ & $0$ & $0$ & $0.4092$ \\ 
			\hline
			$\left\langle P_{e,s}\right\rangle$ & $\frac{\pi}{2}$ & $\frac{\pi}{2}$ & $0.4003$ & $0$ & $\frac{\pi}{2}$ & $0.6487$ \\ 
			\hline
			$\left\langle P_{e,ns}\right\rangle$ & $0$ & $\frac{\pi}{2}$ & $0.3513$ & $0$ & $0$ & $0.5997$ \\ 
			\hline
		\end{tabular}
	}
	\caption{Numerically obtained maximum and minimum values of $\left\langle P_{s}\right\rangle$, $\left\langle P_{ns}\right\rangle$, $\left\langle P_{e,s}\right\rangle$ and $\left\langle P_{e,ns}\right\rangle$ along with the corresponding $\varphi_{1}$ and $\varphi_{2}$ where that maximum or minimum is reached. \label{tb:2}}
\end{table}

\begin{table}[!h]
	\resizebox{0.3\textwidth}{!}{%
		\begin{tabular}{ | c || c | c | c || c | c | c |} 
			\hline
			& $\varphi_{1}$ & $\varphi_{2}$ & min & $\varphi_{1}$ & $\varphi_{2}$ & max \\ 
			\hline \hline
			$P_{abs}$ & $\frac{\pi}{2}$ & $\frac{\pi}{2}$ & $0.1817$ & $0$ & $\frac{\pi}{2}$ & $0.5$ \\ 
			\hline
			$P_{e,abs}$ & $\frac{\pi}{20}$ & $\frac{\pi}{4}$ & $0.0001$ & $0$ & $\frac{\pi}{2}$ & $0.2973$ \\ 
			\hline
		\end{tabular}
	}
	\caption{Numerically obtained maximum and minimum values of $P_{abs}$ and $P_{e,abs}$ along with the corresponding $\varphi_{1}$ and $\varphi_{2}$ where that maximum or minimum is reached. \label{tb:3}}
\end{table}

When noise is considered, the maximum and minimum values in table \ref{tb:2} remain the same in the case when no entanglement is present. However, when entanglement is present, these values change as functions of the parameter $p$ related to the fidelity of the channel. Figure \ref{Pe_noise} shows the dependence of the maximum and minimum values of $\left\langle P_{e,s}\right\rangle$, as a function of the parameter $p$. 

In the classical Monty Hall problem, the host, opening one of the empty boxes, helps the player by creating an imbalance between the probabilities of winning by switching and by not switching, allowing her to make a rational decision. Motivated by this, we define the absolute value of the difference between $\left\langle P_{ns}\right\rangle$ and $\left\langle P_{s}\right\rangle$ for the non-entangled case as

\begin{equation}
P_{abs} = \left| \left\langle P_{ns}\right\rangle - \left\langle P_{s}\right\rangle \right|.
\end{equation}
Analogously, we also define the absolute value of the difference between $\left\langle P_{e,ns}\right\rangle$ and $\left\langle P_{e,s}\right\rangle$ for the entangled case as

\begin{equation}
P_{e,abs} = \left| \left\langle P_{e,ns}\right\rangle - \left\langle P_{e,s}\right\rangle \right|.
\end{equation}

Figure \ref{fig:4} shows the graphics of $P_{abs}$ and $P_{e,abs}$ as functions of $\varphi_{1}$ and $\varphi_{2}$, while table \ref{tb:3} shows their approximate maximum and minimum values as well as the approximate corresponding $\varphi_{1}$ and $\varphi_{2}$ where that maximum or minimum is reached. Notice from this table that it is Bob who can use entanglement to his advantage, either to help or to affect Alice.

\section{Discussion and Conclusions}

In the random approach, when no entanglement is considered, the expectation values of the probabilities of Alice (the player) winning by not switching and by switching, given by equations \eqref{rns} and \eqref{rs} respectively, differ from the classical probabilities by just $0.033$, concluding that, in average, Alice has a better chance of winning by switching, approximately $1.73$ times better, a slightly worse result than in the classical case. When entanglement is considered, as in equations \eqref{rens} and \eqref{res}, the results show the opposite conclusion. In this case, Alice has a better chance of winning by not switching, but just approximately $1.08$ times better, meaning that this kind of correlation between the prize's location and Alice's initial choice, is actually bad for Alice, not allowing her to make a switching choice as meaningful as in the classical case or the quantum non-entangled case.

When Alice and Bob wish to play the random game remotely, noise affects the initial state of the system, modifying just the results where entanglement is considered (figure \ref{PeRan_noise}). We found that, for low values of noise, the switching and not switching gains remain relatively balanced. However, as noise increases, the difference between the gains of the two choices grow too, making the switching decision approximately $2.17$ times better than the not switching one in the worst scenario. This means that, when entanglement is considered, a strongly noisy channel plays in favor of Alice, allowing her to make a rational choice.

In the strategy-based approach, Bob (the host) can freely choose the door-opening parameters, allowing him to increase or decrease (in average) the chances of Alice winning by switching and by not switching. From the results presented in tables \ref{tb:2} and \ref{tb:3}, when entanglement is not considered, we notice that Bob can increase the imbalance between the switching and not-switching cases to a maximum of $0.5$, with $\left\langle P_{s}\right\rangle = 0.75$ and $\left\langle P_{ns}\right\rangle = 0.25$, making the switching decision three times better than the not-switching one, being this the best strategy possible, in average, if Bob wishes to help Alice win the prize. When entanglement is considered and if Bob wants again to help Alice, he can increase the imbalance between the switching and not-switching cases to a maximum of $0.2973$, with $\left\langle P_{e,s}\right\rangle = 0.6487$ and $\left\langle P_{e,ns}\right\rangle = 0.3513$, making the switching decision approximately $1.85$ times better than the not-switching one, a very similar result to the classical case.

As in the random approach, in the strategy-based one, the presence of noise only affects the results where entanglement is considered. In figure \ref{Pe_noise} we see that the switching option becomes dominant for strongly noisy channels, increasing the maximum of the expectation value of the probability to approximately $0.8$. In this scenario, even the minimum value increases to approximately $0.65$, meaning that, as in the random approach, noise plays in favor of Alice.

If Bob does not want Alice to win the prize, his strategy would be to minimize the imbalance between the switching and not-switching cases, decreasing the advantage that she gets from it. Results in tables \ref{tb:2} and \ref{tb:3} show that, when entanglement is not considered, Bob can decrease this imbalance to a minimum of $0.1817$ with $\left\langle P_{s}\right\rangle = 0.5908$ and $\left\langle P_{ns}\right\rangle = 0.4092$, making the switching decision $1.4438$ times better than the not-switching one, still allowing Alice to make a rational choice. However, when entanglement is considered, Bob can decrease the imbalance to a minimum of $0$, as both $\left\langle P_{e,s}\right\rangle$ and $\left\langle P_{ns}\right\rangle$ have the same value of $0.5$ at approximately $\varphi_{1}=\frac{\pi}{20}$ and $\varphi_{2}=\frac{\pi}{4}$, leaving Alice with no other option but to randomly decide whether she bets for switching or not switching.

The proposed experimental set-up, as it is, only allows to model the classical game with three boxes, two players and one empty door to be opened. To extend the set-up to model a more general version of the classical Monty Hall problem as the one discussed in the introduction, there are basically two options: to include more detectors, as these represents the boxes in the game, along with polarization rotators to be able to change the degree of confidence (probability amplitude) in each new box, and to change the two-doors-remain closed condition in order to allow for more doors to be opened. The number of players, however, must remain as two, since the photons produced by the BBO type II crystal come in pairs.

In conclusion, we have presented a quantum version of the Monty Hall problem, based on a quantum-optical set-up that is experimentally feasible. This set-up allows two persons to quickly verify the counter-intuitive result of the famous Monty Hall problem and some statistical results beyond the classical game, adding pedagogical value to the proposed scheme. We have also discussed some results of the game when the parties play through a noisy quantum channel.

\begin{acknowledgments}
L. F. Quezada thanks C3-UNAM for financial support.  A. Mart\'in-Ruiz acknowledges support from DGAPA-UNAM under project No. IA101320. E. Nahmad-Achar acknowledges partial support from DGAPA-UNAM under project No. IN100120.
\end{acknowledgments}

\FloatBarrier

\end{document}